\documentstyle[12pt]{article}
\def\be{\begin{equation}}
\def\ee{\end{equation}}
\def\bc{\begin{center}}
\def\ec{\end{center}}
\def\bea{\begin{eqnarray}}
\def\eea{\end{eqnarray}}

\def\ov{\overline}

\def\mpl{M_{\rm P}}

\catcode`@=11
\def\marginnote#1{}
\newcount\hour
\newcount\minute
\newtoks\amorpm
\hour=\time\divide\hour by60
\minute=\time{\multiply\hour by60 \global\advance\minute by-\hour}
\edef\standardtime{{\ifnum\hour<12 \global\amorpm={am}%
        \else\global\amorpm={pm}\advance\hour by-12 \fi
        \ifnum\hour=0 \hour=12 \fi
        \number\hour:\ifnum\minute<10 0\fi\number\minute\the\amorpm}}
\edef\militarytime{\number\hour:\ifnum\minute<10 0\fi\number\minute}
\def\draftlabel#1{{\@bsphack\if@filesw {\let\thepage\relax
   \xdef\@gtempa{\write\@auxout{\string
      \newlabel{#1}{{\@currentlabel}{\thepage}}}}}\@gtempa
   \if@nobreak \ifvmode\nobreak\fi\fi\fi\@esphack}
        \gdef\@eqnlabel{#1}}
\def\@eqnlabel{}
\def\@vacuum{}
\def\draftmarginnote#1{\marginpar{\raggedright\scriptsize\tt#1}}
\def\draft{\oddsidemargin 0.0truein
        \def\@oddfoot{\sl preliminary draft \hfil
        \rm\thepage\hfil\sl\today\quad\militarytime}
        \let\@evenfoot\@oddfoot \overfullrule 3pt
        \let\label=\draftlabel
        \let\marginnote=\draftmarginnote
   \def\@eqnnum{(\theequation)\rlap{\kern\marginparsep\tt\@eqnlabel}%
\global\let\@eqnlabel\@vacuum}  }
\catcode`@=12
\begin{document}
\begin{titlepage}
\vspace*{-1cm}
\phantom{bla}
\hfill{CERN-TH/95-76}
\\
\phantom{bla}
\hfill{DFPD~95/TH/18}
\\
\phantom{bla}
\hfill{LBL-37021}
\\
\phantom{bla}
\hfill{hep-ph/9504032}
\vskip 1.5cm
\begin{center}
{\Large\bf
Higgs and super-Higgs effects with
\\
naturally vanishing vacuum energy}
\footnote{Work supported in part by the European Union under
contract No.~CHRX-CT92-0004 and by the US DOE under contract
No.DE-AC03-76SF00098.}
\end{center}
\vskip 1.0cm
\begin{center}
{\large Andrea Brignole}\footnote{Supported by an INFN Postdoctoral
Fellowship.} \\
\vskip .1cm
Theory Group, Lawrence Berkeley Laboratory, Berkeley CA 94720, USA
\\
\vskip .2cm
{\large Ferruccio Feruglio}
\\
\vskip .1cm
Dipartimento di Fisica, Universit\`a di Padova, I-35131 Padua, Italy
\\
\vskip .2cm
and
\\
\vskip .2cm
{\large Fabio
Zwirner}\footnote{On leave from INFN, Sezione di Padova, Padua,
Italy.}
\\
\vskip .1cm
Theory Division, CERN, CH-1211 Geneva 23, Switzerland
\end{center}
\vskip 0.5cm
\begin{abstract}
\noindent
We construct $N=1$ supergravity models where the gauge symmetry and
supersymmetry are both spontaneously broken, with naturally vanishing
classical vacuum energy and unsuppressed Goldstino components along
gauge non-singlet directions. We discuss some physically interesting
situations where such a mechanism could play a role, and identify
the breaking of a grand-unified gauge group as the most likely
possibility. We show that, even when the gravitino mass is much
smaller than the scale $m_X$ of gauge symmetry breaking, important
features can be missed if we first naively integrate out the
degrees of freedom of mass ${\cal O} (m_X)$, in the limit of
unbroken supersymmetry, and then describe the super-Higgs effect in
the resulting effective theory. We also comment on possible
connections with extended supergravities and realistic
four-dimensional string constructions.
\end{abstract}
\vfill{
CERN-TH/95-76
\newline
\noindent
March 1995}
\end{titlepage}
\setcounter{footnote}{0}
\vskip2truecm
\vspace{1cm}
{\bf 1.}
If space-time supersymmetry plays a role in the unification of all
fundamental interactions (for a review and references, see e.g.
[\ref{sergio}]), the major obstacle to the construction of
a predictive theory beyond the Standard Model is the problem of
supersymmetry breaking. Whilst useful theoretical tools can be
developed by studying models with global supersymmetry, the only
realistic framework for the discussion of such a problem is $N=1$
supergravity, regarded as the low-energy limit of a consistent
quantum theory including gravity.

In supergravity, gravitational interactions are {\em always} relevant
in the discussion of the super-Higgs phenomenon, and we must face
the highly non-trivial requirement of a sufficiently small
cosmological constant. In this respect, promising starting points
are the $N=1$ supergravity models characterized by a
positive--semi-definite classical potential, with all minima
corresponding to broken supersymmetry and vanishing vacuum energy,
and the gravitino mass sliding along some flat direction
[\ref{noscale},\ref{lhc}].

Along this line of thought, attention has mainly concentrated on the
case in which both the Goldstino field and the flat directions are
singlets under the full gauge group. Only recently, the possibility
was considered of breaking supersymmetry and $SU(2) \times U(1)$
at once, with naturally vanishing vacuum energy [\ref{bz}]: an
explicit model of this kind was produced, but the gauge and Yukawa
interactions of the Goldstino were suppressed down to gravitational
strength, ${\cal O}(m_{3/2}/\mpl)$, by mixing effects involving
some singlet moduli fields.

In this paper, we examine the possibility of breaking the gauge
symmetry together with supersymmetry, with a naturally vanishing
classical vacuum energy and unsuppressed Goldstino components
along gauge non-singlet directions. In section~2, we present a
toy model that provides an existence proof for this possibility
and allows a number of issues of general relevance to be discussed
in a simplified setting. In section~3, we discuss how our results
could be extended to more realistic situations: the breaking
of the electroweak symmetry, of a grand-unified symmetry, or
of a gauge symmetry of a strongly interacting hidden sector.
In section~4, we conclude with some comments on the possible
connections with extended supergravities and four-dimensional
string models. To improve the readability of the text, we have
collected some useful formulae in an Appendix.

\vspace{1cm}
{\bf 2.}
Consider an $N=1$ supergravity model containing three chiral
superfields, whose complex spin~0 components parametrize the
K\"ahler manifold:
\be
\label{katoy}
\left[ {SU(1,1) \over U(1)} \right]_S \times \left[ {SO(2,2) \over
SO(2) \times SO(2)} \right]_{T,U} \simeq \left[ {SU(1,1) \over
U(1)} \right]^3_{S,T,U} \, .
\ee
The K\"ahler potential can be conveniently written as\footnote{Unless
otherwise stated, we use the standard supergravity conventions where
$\mpl \equiv 1/\sqrt{8 \pi G_N} = 1$.}
\be
K = - \log Y \, ,
\ee
where, using one of the parametrizations of $SO(2,2)/ [SO(2) \times
SO(2)] \simeq [SU(1,1) / U(1)]^2$ discussed in the Appendix,
\be
\label{prima}
Y = \left( S + \ov{S} \right)
\left( T + \ov{T} \right) \left( U + \ov{U} \right) \, .
\ee
This parametrization has the advantage that a constant
superpotential, $w = k \ne 0$ (where it is not restrictive
to choose $k$ real and positive), gives an identically
vanishing classical potential with a non-vanishing
gravitino mass
\be
e^{\cal G} = { k^2 \over
\left( S + \ov{S} \right)
\left( T + \ov{T} \right)
\left( U + \ov{U} \right) }
\, .
\ee
On the other hand, if one sticks to this parametrization one cannot
introduce any gauge symmetry acting linearly but non-trivially on the
fields.

As discussed in the Appendix, one can move to an alternative
parametrization,
\be
\label{seconda}
Y = \left( S + \ov{S} \right)
\left(1 - |H_1|^2 \right) \left( 1 - |H_2|^2 \right) \, .
\ee
The constant superpotential $w=k$ would now become
\be
\label{wuno}
w = {k \over 2}  (1+H_1)  (1+H_2) \, .
\ee
The K\"ahler potential corresponding to eq.~(\ref{seconda})
is invariant under two continuous $U(1)$ groups, whose
generators will be denoted by $X_1$ and $X_2$, acting
linearly but non-trivially on the fields $H_1$ and $H_2$.
One could think of gauging some non-anomalous combination of
them, but such an attempt must face the fact that the superpotential
of eq.~(\ref{wuno}) would explicitly break gauge-invariance.

A possible way out is to replace the superpotential of
eq.~(\ref{wuno}) by
\be
\label{wtoy}
w = k \left( 1 + \sqrt{H_1 H_2} \right)^2 \, .
\ee
The ambiguity in the relative phase between the two terms
within brackets can be removed by a phase redefinition of the
$H_{1,2}$ fields.
The model defined by eqs.~(\ref{seconda}) and (\ref{wtoy})
admits the gauge group $G_0=U(1)_X$ if one assigns to
$(S,H_1,H_2)$ the (arbitrarily normalized) charges,
corresponding to the combination $X \equiv X_1 - X_2$:
\be
\label{toycharges}
X(S)=0 \, ,
\;\;\;\;\;
X(H_1)=-1/2 \, ,
\;\;\;\;\;
X(H_2)=+1/2 \, .
\ee
We fix the arbitrariness in the choice of the gauge kinetic function
by taking, for the time being, $f=S$ (alternative choices will be
discussed at the end of this section). Then a well-behaved gauge
coupling and K\"ahler
metric require $s \equiv S + \ov{S} > 0$ and either $|H_1|,|H_2|<1$
or $|H_1|,|H_2|>1$. Moreover, analyticity of the superpotential
excludes from the acceptable field configurations the lines $H_1=0$
and $H_2=0$. The continuous $[SU(1,1)]^3$ symmetry of the K\"ahler
manifold is explicitly broken by the superpotential $w$ and by the
gauge kinetic function $f$, with the exception of the $U(1)_X$ gauge
symmetry. It is also interesting to notice that the discrete
transformations $(H_1 \rightarrow 1/H_1, H_2 \rightarrow 1/H_2)$
and $(H_1 \rightarrow H_2, H_2 \rightarrow H_1)$, which are not
contained in $[SU(1,1)]^3$, are also symmetries of the model.

The full classical potential for our model reads $V_0 = V_F + V_D$,
where
\be
\label{vftoy}
V_F = {k^2 \over s} {(|H_1|-|H_2|)^2 (1 + |H_1| |H_2|) | 1 +
\sqrt{H_1 H_2}|^2 \over |H_1| |H_2| (1 - |H_1|^2) (1 - |H_2|^2)}
\, ,
\ee
\be
\label{vdtoy}
V_D = {1 \over 4 s} {(|H_1|^2-|H_2|^2)^2 \over (1 - |H_1|^2)^2
(1 - |H_2|^2)^2}
\, .
\ee
It is easy to see that $V_0$ is positive semi-definite, and admits a
continuum of degenerate minima with broken gauge symmetry, broken
supersymmetry and vanishing vacuum energy, corresponding to
arbitrary $|H_1| = |H_2|$ and $S$. It may be useful to reinterpret
these flat directions in terms of continuous symmetries of the
classical potential and of its minimization conditions. The only
continuous symmetry of $V_0$, besides the gauged $U(1)_X$, is the
non-compact $U(1)$ corresponding to imaginary translations of the
$S$ field [the $U(1)_{\hat{X}}$ associated with $\hat{X} \equiv
X_1 + X_2$ is an invariance of $V_D$ but not of $V_F$]. However,
the minimization conditions $V_F=V_D=0$ defining the classical
vacua inherit as symmetries the full $[SU(1,1)]_S$, and common
phase rotations and rescalings of $H_1$ and $H_2$, corresponding
to the complexification of $U(1)_{\hat{X}}$: we then expect four
massless real spin-0 degrees of freedom, besides the would-be
Goldstone boson of the broken $U(1)_X$.

In order to examine the classical moduli space of our theory, we
recall that there are in principle three independent gauge-invariant
VEVs, $|H_1|$, $|H_2|$ and $\theta \equiv {\rm arg} \, (H_1 H_2)$.
The minimization condition $V_0=0$ requires $|H_1|=|H_2| \equiv h$,
so we need $h$ and $\theta$ to label the physically inequivalent
vacua (apart from the residual redundancy due to the unbroken
discrete symmetries).

The order parameters for gauge and supersymmetry breaking are the
vector boson mass
\be
\label{mvectoy}
m_X^2 = {2 h^2 \over s (1-h^2)^2} \, ,
\ee
and the gravitino mass
\be
\label{mgratoy}
m_{3/2}^2 = {k^2 |1+e^{i \theta / 2} h|^4 \over s (1-h^2)^2} \, ,
\ee
respectively. In the spin-0 sector, the only massive state
corresponds to ${\rm Re} \, [ (H_1 - H_2) e^{-i \theta /2}]$,
with mass
\be
\label{mscatoy}
m_0^2 = m_X^2 + m_{3/2}^2 {2 (1 + h^2) (1-h^2)^2 \over
h^2 |1+ h e^{i \theta/2}|^2 } \, .
\ee

In the spin-1/2 sector, a $4 \times 4$ mass matrix describes the
mixing of the fields $\tilde{S}$, $\tilde{H_1}$, $\tilde{H_2}$
and $\tilde{X}$ (gaugino), which here are understood to be
already canonically normalized. It is particularly convenient to
introduce the symmetric and antisymmetric higgsino combinations,
$\tilde{H}_S \equiv (\tilde{H_1} + \tilde{H}_2)/\sqrt{2}$ and
$\tilde{H}_A \equiv (\tilde{H_1} - \tilde{H}_2)/\sqrt{2}$, because
the $4 \times 4$ fermionic mass matrix can then be decomposed into
two $2 \times 2$ blocks, one for $(\tilde{S},\tilde{H}_S)$ and the
other for $(\tilde{H}_A, \tilde{X})$. The first block has a
vanishing eigenvalue corresponding to the Goldstino and another
eigenvalue $m_1^2 = m_{3/2}^2$. The (canonically normalized)
Goldstino can be written as $\tilde{\eta} = (\tilde{S} + \sqrt{2}
e^{i \phi} \tilde{H}_S)/\sqrt{3}$, where $e^{i \phi} = (1 +
e^{i\theta/2}h)/(1 + e^{-i\theta/2}h)$. Notice that, since the
Goldstino component along $\tilde{H}_S$ is unsuppressed, we obtain
a gravitino with interactions of gauge strength via its $\pm 1/2$
helicity components [\ref{fayet}]. The second block has eigenvalues
\be
\label{mfertoy}
m_{2,3}^2 = m_X^2 + m_{3/2}^2  \left[  1 + { (1 + h^2) (1-h^2)^2
\over 2 h^2 |1+ h e^{i \theta/2}|^2 } \right]
\pm {1 + h^2 \over h} m_{3/2} \sqrt{ m_{3/2}^2
{(1 - h^2)^4 \over 4 h^2 |1+ h e^{i \theta/2}|^4 } + m_X^2} \, .
\ee

Observe that, in the model under consideration,
\be
\label{strtoy}
{\rm Str \;} {\cal M}^2 (h,\theta,S) \equiv \sum_i (-1)^{2 J_i}
(2 J_i + 1) m_i^2 (h,\theta,S) = - 10 m_{3/2}^2  \, ,
\ee
where the only dependence on the variables $h$, $\theta$ and
$S$ is the implicit one through the gravitino mass. Such a
property is phenomenologically welcome, since it may allow
for a natural cancellation of the quadratically divergent
quantum corrections to the vacuum energy from other
sectors of the theory [\ref{lhc}]. For example, we could add
$n$ chiral superfields $z$, with canonical kinetic terms and
superpotential at least quadratic in $z$: in this case we
would obtain, around the minima with $z=0$, an additional
contribution $\Delta \; {\rm Str \;} {\cal M}^2 = 2 \, n \,
m_{3/2}^2$.

We would like to stress that, as expected, the mass spectrum is
invariant under the discrete transformation $h \rightarrow 1/h,
\theta \rightarrow - \theta$, so it will not be restrictive to
study it for $0 < h < 1$.

Some interesting limits of our model are $h \to 0$ (equivalent to
$h \to \infty$) and $h \to 1$ (with $\theta$, $k$ and $s$ fixed).

For $h \to 0$ we obtain $m_X^2 \to 0$, $m_{3/2}^2 \to k^2/s$,
i.e. unbroken gauge symmetry with broken supersymmetry\footnote{
Notice that the superpotential $w$, restricted to the classical
moduli space, is singular at $h=0$, with monodromy $(H_1 \to - H_1,
H_2 \to - H_2)$, corresponding to $(T \to 1/T, U \to 1/U)$ in the
alternative parametrization.}. Observe
that, in this limit, $m_0^2 \to 2 m_{3/2}^2 / h^2 + \ldots$,
$m_2^2 \to m_{3/2}^2 / h^2 + \ldots$, where the dots stand for
terms that do not diverge in the limit. Reintroducing
explicitly the Planck mass for clarity, for $h \ll \sqrt{m_{3/2}
\mpl}$ there are supersymmetry-breaking mass splittings $\Delta
m^2$ much larger than $m_{3/2}^2$, with couplings of order
$\Delta m^2 / ( m_{3/2} M_P) \sim m_{3/2} M_P / h^2 \gg 1$,
and we end up with a strongly interacting Goldstino.

For any fixed $\theta \ne 2 \pi$, the limit $h \to 1$ corresponds
(formally) to maximally broken gauge symmetry and supersymmetry,
i.e. $m_X^2 , m_{3/2}^2 \to \infty$. This is not the case for
the special value $\theta = 2 \pi$, for which $h \to 1$ corresponds
to $m_X^2 \to \infty$ but $m_{3/2}^2 \to 0$.

It may be useful to rephrase the previous results in the
alternative $(T,U)$ parametrization: the classical vacua
correspond to $T=U$, and the singular points to $T=U=1$ and $T +
\ov{T} = U + \ov{U} = 0, \infty$.

Another physically interesting limit is the case in which
$m_{3/2} \ll m_X$ (which is realized, for example, for $k \ll 1$
and $h$ generic). In this case one can write down a low-energy
effective field theory for the light modes. Such a theory has no
residual gauge symmetry, and its chiral supermultiplet content
consists only of $S$ and $H_S \equiv (H_1 + H_2)/\sqrt{2}$. After
an innocuous rescaling $H_S \to \sqrt{2} H_S$, its K\"ahler
potential and superpotential are given by
\be
\label{effth}
Y = (S + \ov{S}) (1 - |H_S|^2)^2 \, ,
\;\;\;\;\;
w = k (1 + H_S)^2 \, ,
\ee
and give an identically vanishing classical potential. Notice
that this effective theory cannot correctly reproduce the
singular behaviour of the full theory for $h \to 0$: as trivial
as it sounds, this may be interpreted as a warning for the
discussion of modular covariant superpotentials in superstrings
effective supergravities.  Notice also that, when $m_{3/2} \ll
m_X$, in the effective theory below the scale $m_X$ we would find
${\rm Str \;} {\cal M}^2 = - 6 m_{3/2}^2$, which differs from
eq.~(\ref{strtoy}): this is just reminding us that ${\rm Str \;}
{\cal M}^2$ is a physically meaningful object, in relation with
the stability of the flat background and of possible gauge
hierarchies, only when computed over all states of the fundamental
theory that get supersymmetry-breaking mass splittings.

One could also consider more complicated limits involving
combinations of $k$, $s$, $h$ and $\theta$, but we shall
not pursue this type of considerations further.

Before leaving our toy model for the discussion of more realistic
situations, we would like to comment on some possible variants.
One may ask if there are forms of the gauge kinetic function $f$,
more general than $f=S$, that respect gauge invariance and
allow for ${\rm Str \;} {\cal M}^2 = ({\rm constant}) m_{3/2}^2$.
On the vacua with $\theta=0$ and $S$ real, a class of functions
satisfying this requirement is
\be
\label{fgen}
f = \left( S {1 - \sqrt{H_1 H_2} \over 1 + \sqrt{H_1 H_2}}
\right)^{\displaystyle -c/2} \cdot \varphi \left( S {1 + \sqrt{H_1
H_2}
\over 1 - \sqrt{H_1 H_2}} \right) \, ,
\ee
where $c$ is an arbitrary real constant and $\varphi(z)$ is an
arbitrary holomorphic function. The original choice $f=S$ is
recovered for $c=-1$ and $\varphi(z)=\sqrt{z}$. For the general
gauge kinetic function of eq.~(\ref{fgen}), the supertrace
formula of eq.~(\ref{strtoy}) becomes
\be
\label{strtoygen}
{\rm Str \;} {\cal M}^2 (h,\theta=0,S)
= - 2 (4 + c^2) m_{3/2}^2  \, .
\ee
As a curiosity, observe that, choosing $\varphi(z) = z^{c/2}$, we
get $f=[(1+\sqrt{H_1H_2})/(1-\sqrt{H_1H_2})]^c$. The transformation
$(H_1 \to - H_1, H_2 \to - H_2)$, associated with the monodromy of
$w$ around $h=0$, would correspond in this case to a weak/strong
coupling duality $f \to 1/f$.

Another possibility is to look for different gaugings of the
sigma model under consideration. For example, one could make
the additional field redefinition $S=(1-z)/(1+z)$, and introduce
the superpotential $w = k [ 1 + (zH_1H_2)^{1/3} ]^3$. This would
allow two independent $U(1)$ factors to be gauged, producing a
positive--semi-definite potential, broken supersymmetry at all
classical vacua, and less flat directions than in the model defined
by eqs.~(\ref{seconda}) and (\ref{wtoy}). As a candidate form for
the gauge kinetic function $f_{ab}$ ($a,b=1,2$), it is interesting
to consider in this case $f_{ab}= k_a \delta_{ab} \{ [ 1 +
(zH_1H_2)^{1/3}] / [ 1 - (zH_1H_2)^{1/3}]\}^r$, which gives, on the
vacua with $z=H_1=H_2  \in {\bf R}^+$, a gaugino mass $m_{1/2} = r
\, m_{3/2}$, and has also interesting properties with respect to
weak/strong coupling duality.

Yet another variant would consist in removing the $S$ field
(either explicitly or by introducing a superpotential that
gives a VEV to its scalar component without giving a VEV to
its auxiliary component), and in assigning to the fields $(H_1
,H_2)$ the K\"ahler potential $K = - (3/2) \log [ ( 1 - |H_1|^2)( 1
- |H_2|^2)]$ and the superpotential $w=k ( 1 + \sqrt{H_1 H_2})^3$.
Choosing $f= L [(1 + \sqrt{H_1 H_2})/(1 - \sqrt{H_1 H_2})]^c$,
with $L$ arbitrary constant and $c \in {\bf R}$, would give a gaugino
mass $m_{1/2}= c \, m_{3/2}$ at all minima with $H_1=H_2 \in {\bf
R}$;
the choice $c = \pm 1$ and $L \in {\bf R}$ would guarantee $m_{1/2}^2
= m_{3/2}^2$ at all minima, corresponding to $|H_1|=|H_2|$, but would
break the discrete invariance under $(H_1 \to 1/H_1, H_2 \to 1 /
H_2)$.

\vspace{1cm}
{\bf 3.}
Supergravity models of the type considered in the previous section,
with gauge symmetry and $N=1$ supersymmetry both spontaneously
broken, and naturally vanishing classical vacuum energy, can be
obtained by the following procedure. First, one selects a K\"ahler
manifold for the symmetry-breaking sector. For K\"ahler manifolds
of the type $G/H$, where $H$ is the maximal compact subgroup of $G$,
one chooses the gauge group $G_0$ as a subgroup of $H$ (this can be
obviously generalized to a factorized manifold of the type $G/H
\times
M$, where $M$ is a sub-manifold parametrized by some gauge-singlet
fields). To ensure manifest gauge-invariance, it is convenient to
work in a parametrization of $G/H$ where $H$ is linearly realized.
For example, in the case where $G=SU(m,n)$ and $H = SU(m) \times
SU(n) \times U(1)$, the scalar fields can be described by an $m
\times n$ complex matrix $Z$, with the K\"ahler potential for $G/H$
given by [\ref{cdf}]
\be
\label{kasu}
K = - \log \det \left( 1 - Z Z^{\dagger} \right) \, ,
\ee
where $1$ denotes the unit $m \times m$ matrix; $K$ is manifestly
invariant under the transformations
\be
\label{transfsu}
Z' = e^{i \alpha} U Z V^{\dagger} \, ,
\ee
where $\alpha$ is a real parameter, and $U$ and $V$ are $SU(m)$ and
$SU(n)$ matrices, respectively. Another important example, which
appears in the effective supergravity theories of many
four-dimensional
string constructions, corresponds to $G=SO(2,n)$ and $H = SO(2)
\times
SO(n)$. The scalar fields are described by the $n$-dimensional
complex
vector $y$. The K\"ahler potential reads [\ref{cv}]
\be
\label{kaso}
K = - \log \det \left( 1 - 2 y^{\dagger} y + | y^T y |^2 \right) \, ,
\ee
and is manifestly invariant under the transformations
\be
\label{transfso}
y' = e^{i \alpha} O y \, ,
\ee
where $\alpha$ is a real parameter and $O$ is an $SO(n)$ matrix.
In the parametrizations specified by eqs.~(\ref{kasu}) and
(\ref{kaso}), the full $H$ subgroup of $G$ is linearly realized
and the K\"ahler potential is strictly gauge-invariant. One then
looks for a gauge-invariant superpotential $w$ that breaks
simultaneously supersymmetry and the gauge symmetry with naturally
vanishing vacuum energy. Needless to say, additional physical
criteria can be used to constrain the possible forms of the
superpotential: we have in mind, for example, generalized
duality symmetries and singularity structure of strings effective
supergravities (for a review and references see e.g. [\ref{gpr}]).
One can then couple additional sectors of the theory,
which do not take part in the symmetry breaking mechanism, by
specifying their contributions to the K\"ahler potential and to the
superpotential.

We now discuss some physically relevant situations where the
general mechanism discussed above may be at work.

The first possibility that comes to mind is to associate the
breaking of supersymmetry with the breaking of the electroweak
gauge symmetry, $SU(2)_L \times U(1)_Y$, in the Minimal
Supersymmetric extension of the Standard Model (MSSM) coupled
to $N=1$ supergravity. Since the MSSM Higgs sector contains
the two doublets $H_1 \equiv (H_1^0,H_1^-)$ and $H_2 \equiv
(H_2^+,H_2^0)$, a natural choice is to consider the K\"ahler
manifold
\be
\label{kamssm}
{SO(2,4) \over SO(2) \times SO(4)}
\simeq
{SU(2,2) \over SU(2) \times SU(2) \times U(1)}
\, ,
\ee
parametrized by the $2 \times 2$ complex matrix\footnote{An
analogous description, in the absence of gravity, can be found
in [\ref{fmp}].}
\be
Z \equiv \left( \begin{array}{cc}
H_1^0 & H_2^+ \\ H_1^- & H_2^0 \end{array} \right) \, ,
\ee
with the K\"ahler potential of eq.~(\ref{kasu}). To recover the
usual $SU(2)_L \times U(1)_Y$ gauge transformations of the doublets
$H_1$ and $H_2$, with parameters $\alpha_A$ ($A=1,2,3$) and
$\alpha_Y$, one must consider eq.~(\ref{transfsu}) with $\alpha=0$,
$U= \exp ( i \alpha_A \tau^A/2)$, and $V = \exp ( i \alpha_Y
\tau^3/2)$. Inspired by
the structure of string effective supergravities and by the analogy
with our toy model, we also introduce a singlet field $S$,
parametrizing
a factorized $SU(1,1)/U(1)$ manifold; we assume a gauge-invariant
superpotential of the form
\be
\label{wmssm}
w_0 = k \left( 1 + \sqrt{\det Z} \right)^2 \, ,
\ee
which represents the obvious generalization of the one of
eq.~(\ref{wtoy}).
This leads to a positive--semi-definite tree-level potential,
identically vanishing for arbitrary $s$ and
\be
\label{zvacua}
Z = h e^{i \theta / 2} A \, .
\ee
In eq.~(\ref{zvacua}), $h$ and $\theta$ are arbitrary real numbers,
and $A$ is an arbitrary $SU(2)$ matrix, which can be reabsorbed by
an $SU(2)_L \times U(1)_Y$ gauge transformation. Thus the classical
moduli space in the $Z$ sector, describing the broken phase in which
only $U(1)_{em}$ survives, can be parametrized in terms of $h$ and
$\theta$. With the choice $f_{ab} = \delta_{ab} S$ ($g^2 = g'^2 =
2 / s$), the spectrum is an obvious generalization of the
toy-model one. Notice that, for ${\cal O} (1)$ gauge couplings,
to obtain $m_{W,Z}/\mpl \sim 10^{-16}$ one must choose $h/\mpl
\sim 10^{-16}$, leading to $m_{3/2}/\mpl \sim k / \sqrt{s}$. The
tree-level supersymmetry-breaking mass splittings in the gauge-Higgs
sector are either vanishing or of order $\Delta m^2 \sim m_{3/2}^2
\mpl^2 / h^2$. Thus for $m_{3/2} \sim h^2 / \mpl \sim 10^{-4} {\; \rm
eV}$ the non-vanishing splittings are of order $h$ and the Goldstino
couplings are of order unity, whereas for $m_{3/2} \sim h$ one gets
non-vanishing splittings of order $\mpl$ and Goldstino couplings
outside the perturbative regime.

To complete the model, one should also specify the K\"ahler potential
and the superpotential involving the quark and lepton superfields
$z$.
If $m_{3/2} \sim h^2 / \mpl$, one has to face the same problem as in
the models with spontaneously broken global supersymmetry: one
typically obtains at least one squark of charge $1/3$ lighter than
the corresponding quark [\ref{dg}], which
is excluded by the present experimental bounds. If $m_{3/2} \sim h$,
one can obtain an acceptable spectrum of squarks and sleptons,
for example choosing canonical kinetic terms and a superpotential
\be
w = w_0 \left( 1 + {h^U Q U^c H_2 + h^D Q D^c H_1 + h^E L E^c H_1
\over \sqrt{\det Z}} \right) \, ,
\ee
where $w_0$ is the superpotential of eq.~(\ref{wmssm}). However,
the presence of huge mass splittings of order $\mpl$ in the
gauge-Higgs sector, associated with non-perturbative Goldstino
couplings, does not allow us to control the quantum corrections.
If we naively compute the one-loop corrections to the effective
potential, imagining $m_{3/2}$ fixed and considering only the
leading $h$-dependence, we find that the
classical degeneracy is removed to give $h \sim m_{3/2}$ at the
one-loop minimum, but we cannot trust this result in the absence of
tools to control higher order and non-perturbative effects.

In summary, the structure discussed for the toy model does not seem
suitable for a direct application to $SU(2)_L \times U(1)_Y$
breaking. For a more satisfactory description of the latter, one
may be forced to introduce some extra $G_{SM}$-singlets as in
ref.~[\ref{bz}].

A second, more intriguing possibility is to associate the breaking of
supersymmetry with the breaking of a grand-unified gauge
group\footnote{The combined breaking
of supersymmetry and of a grand-unified gauge symmetry
was previously considered in [\ref{mariano}], but the vanishing
of the classical vacuum energy was achieved there by fine-tuning
some superpotential parameters.} $G_U$ down to the MSSM gauge group
$G_{SM} \equiv SU(3)_C \times SU(2)_L \times U(1)_Y$. Various
realizations could be possible, depending on the choice of $G_U$ and
of the K\"ahler manifold for the Higgs sector. We do not commit
ourselves here to any specific example, but we just use the toy model
as a guideline for a qualitative discussion. Given the approximate
phenomenological relation $M_U \sim g_U \mpl$, also suggested by
four-dimensional string models, we need $h$ to be of
order $\mpl$. Assuming as before $f=S$, supersymmetry-breaking mass
splittings will then be of order $m_{3/2}$, signalling a Goldstino
with interactions of gravitational strength if we take $m_{3/2}$ at
the electroweak scale as usual. In this case a perturbative study of
the dynamical determination of $M_U$ and $m_{3/2}$ could be possible,
and one may also find applications to the doublet-triplet splitting
problem.

The previous list does not exhaust the physically interesting
possibilities. For example, one may imagine a strongly interacting
hidden sector where non-perturbative phenomena break supersymmetry
as well as the gauge symmetry $G_{hid}$ down to a subgroup $H_{hid}$.

\vspace{1cm}
{\bf 4.}
The new class of supergravity models discussed in the previous
sections has in our opinion rather intriguing properties (including
some formal similarities with recent and less recent results results
on non-perturbative phenomena in globally supersymmetric theories
[\ref{npglo}]), but suffers from two main unsatisfactory aspects. The
first is connected with the apparent arbitrariness of the
construction:
at the level of $N=1$ supergravity, we are practically free to choose
the gauge group, the number of chiral superfields, the K\"ahler
manifold, the embedding of the gauge group in the isometry group of
the K\"ahler manifold, and finally the gauge kinetic function and
the superpotential that breaks supersymmetry. The second is connected
with the fact that, at the level of $N=1$ supergravity, we are
essentially bound to a classical treatment, given the ambiguities
of an effective, non-renormalizable theory in the control of quantum
corrections, both perturbative and non-perturbative. One may hope to
improve in both directions by establishing some connections with
extended $N>1$ supergravity theories and especially with
four-dimensional
superstring models.

To obtain a realistic $N=1$ supergravity model, only the candidate
quark and lepton superfields need to transform in chiral
representations
of the gauge group. It is then conceivable that the sector involved
in the Higgs and super-Higgs effects can be obtained, by some
suitable
projection, from the gauge and gravitational sectors of an extended
supergravity model. Indeed, spontaneous supersymmetry breaking with
vanishing classical vacuum energy can be associated, in $N=2$
[\ref{neq2}], $N=4$ [\ref{neq4}] and $N=8$ [\ref{neq8}] supergravity,
with the gauging of a non-compact subgroup of the duality group.
The examples we are aware of give gauge-singlet Goldstinos in the
resulting $N=1$ theory, but one could look for models where the
projected $N=1$ Goldstino transforms non-trivially under the $N=1$
gauge group: such models would satisfy highly non-trivial
constraints, due to the underlying extended supersymmetry.

Further constraints could be obtained by deriving models of the type
discussed in this paper as low-energy effective theories of
four-dimensional string models with spontaneously broken
$N=1$ supersymmetry. This looks like a natural possibility:
we know many examples of singlet moduli appearing in the
effective string supergravities that are indeed flat directions
breaking an underlying gauge group, restored only at points of
extended symmetry. Unfortunately, the only existing examples
[\ref{ss}] are those in which supersymmetry is broken at the
string tree-level, via coordinate-dependent orbifold
compactifications, and correspond to cases where the Goldstino
direction is a gauge singlet. It should be possible to extend
these constructions to models where the gauge symmetry and
supersymmetry are both spontaneously broken. This could lead to
some progress in the control of perturbative quantum corrections,
since, working at the string level and not in the effective field
theory, one can compute the full spectrum of states that contribute
to the one-loop partition function.

However, the previous approach looks hopeless as far as the
dynamical determination of the dilaton VEV is concerned,
since the latter must involve some non-perturbative mechanism.
Still, one could use the knowledge of some effective string
supergravities in the limit of unbroken supersymmetry, even in the
version
including infinitely many lattice states [\ref{gp}], and parametrize
possible non-perturbative effects with suitable modifications of the
superpotential and of the gauge kinetic function, respecting the
quantum symmetries of the underlying string theory.

One could take, as a modest but concrete example, one of the
$N=1$ four-dimensional fermionic string constructions [\ref{ferstr}]
that give gauge groups such as $SO(10) \times \ldots$ or flipped
$SU(5) \times U(1) \times \ldots$ [\ref{flipped}], where the
dots stand for some hidden-sector gauge group. In the limit of
unbroken
$N=1$ supersymmetry, their classical effective theories are known
[\ref{fereff}]. One could then look for gauge-invariant
superpotential
modifications that break  the gauge symmetry down to $G_{SM}$ and
supersymmetry at the same time, with naturally vanishing vacuum
energy.

We hope to return to these problems in a future publication.

\vfill{
\section*{Acknowledgements}
We would like to thank L.~Alvarez-Gaum\'e, S.~Dimopoulos,
L.E.~Ib\'a\~nez, and especially  S.~Ferrara and C.~Kounnas,
for useful discussions. The work of A.B. is supported in
part by the US DOE under contract No.DE-AC03-76SF00098.
}

\newpage
\section*{Appendix}

We collect here some useful formulae for the K\"ahler manifold
\be
{SO(2,2) \over SO(2) \times SO(2)}
\simeq
\left[ {SU(1,1) \over U(1)} \right]^2
\, .
\ee
Its K\"ahler potential can be written in general as [\ref{fereff}]
\be
K = - \log Y \, ,
\;\;\;\;\;
Y = 2 \left( |a|^2 + |b|^2 - | \varphi_1 |^2 -
| \varphi_2 |^2 \right) \, ,
\ee
where $(a,b,\varphi_1,\varphi_2)$ are analytic functions of
two unconstrained complex fields, satisfying the requirement
\be
a^2 + b^2 - \varphi_1^2 - \varphi_2^2 = 0 \, .
\ee
One useful parametrization is
\be
\label{first}
Y^{(i)} = \left( T + \ov{T} \right)
\left( U + \ov{U} \right) \, ,
\ee
corresponding to $a=(1+UT)/2$, $b=i(U+T)/2$,
$\varphi_1=(1-UT)/2$, $\varphi_2=i(U-T)/2$.
In the parametrization of eq.~(\ref{first}),
each of the two $SU(1,1)$ factors acts as follows:
\be
T \longrightarrow {a T - i b \over i c T + d}
\;\;\;\;\;
(ad - bc = 1) \, ,
\ee
and modifies the K\"ahler potential by a K\"ahler transformation
\be
T + \ov{T} \longrightarrow {T + \ov{T}  \over |i c T + d|^2} \, .
\ee
In particular, $Y^{(i)}$ is strictly invariant under the
continuous $U(1)$ associated with imaginary translations,
$T \rightarrow T - i b$. Similar relations can be obtained
for the $U$ field. Another continuous invariance of the
K\"ahler potential corresponds to the rescalings
\be
\label{rescalings}
T \rightarrow \lambda T \, ,
\;\;\;\;\;
U \rightarrow {1 \over \lambda} U \, .
\ee
Finally, $Y^{(i)}$ is strictly invariant under some additional
discrete transformations that do not belong to $[SU(1,1)]^2$:
\be
\label{discrone}
T \rightarrow - T \, ,
\;\;
U \rightarrow - U \, ,
\ee
\be
\label{discrtwo}
T \rightarrow U \, ,
\;\;
U \rightarrow T \, .
\ee
In the parametrization of eq.~(\ref{first}), the K\"ahler
potential is well defined in the two domains
\be
\label{domone}
(T + \ov{T})  , (U + \ov{U}) > 0  \, ,
\;\;\;\;\;
(T + \ov{T}) , (U + \ov{U}) < 0  \, .
\ee

A second useful parametrization is
\be
\label{second}
Y^{(ii)} = \left( 1 - \left| H_1 \right|^2 \right)
\left( 1 - \left| H_2 \right|^2 \right) \, ,
\ee
corresponding to $a=(1+H_1H_2)/2$, $b=i(1-H_1H_2)/2$,
$\varphi_1=-(H_1+H_2)/2$, $\varphi_2=i(H_1-H_2)/2$.
Equations~(\ref{first}) and (\ref{second}) are connected by
the field redefinitions $T=(1-H_1)/(1+H_1)$ $[H_1=(1-T)
/(1+T)]$, and similarly for $U$ and $H_2$. Notice that
the two K\"ahler potentials are equivalent only up to a
K\"ahler transformation, corresponding to a multiplicative
superpotential modification:
\be
T + \ov{T} = {1 - |H_1|^2 \over \left|
{1 + H_1 \over \sqrt{2}} \right|^2} \, ,
\;\;\;\;\;
1 - |H_1|^2 = { T + \ov{T} \over \left|
{1 + T \over \sqrt{2}} \right|^2} \, ,
\ee
and similarly for $U$ and $H_2$. This is consistent
with the fact that $(a,b,\varphi_1,\varphi_2)$ are defined
only up to a universal multiplicative function of the
unconstrained fields. In the parametrization of eq.~(\ref{second}),
each of the $SU(1,1)$ factors acts as follows
\be
H_1 \longrightarrow {\xi H_1 + \eta \over \ov{\eta} H_1 + \ov{\xi}}
\, , \;\;\;\;\; ( |\xi|^2 - |\eta|^2 = 1) \, ,
\ee
\be
\xi = { (d + a) + i (b-c) \over 2} \, ,
\;\;\;\;\;
\eta = { (d - a) + i (b+c) \over 2} \, ,
\ee
and modifies the K\"ahler potential by a K\"ahler transformation
\be
1 - |H_1|^2 \longrightarrow {1 - |H_1|^2 \over |\ov{\eta} H_1 +
\ov{\xi}|^2 } \, .
\ee
In particular, the K\"ahler potential is strictly invariant under
the continuous $U(1)$ associated with phase rotations, $H_1
\rightarrow
e^{i \theta} H_1$. Similar relations can be obtained for the $H_2$
field. The continuous invariance of eq.~(\ref{rescalings})
is not realized in a simple form. On the other hand, the discrete
invariances of eqs.~(\ref{discrone}) and (\ref{discrtwo}) take the
suggestive forms
\be
\label{discruno}
H_1 \rightarrow {1 \over H_1} \, ,
\;\;
H_2 \rightarrow {1 \over H_2} \, ,
\ee
\be
\label{discrdue}
H_1 \rightarrow H_2 \, ,
\;\;
H_2 \rightarrow H_1 \, .
\ee
The two domains in which the K\"ahler potential is well defined
become
\be
\label{domuno}
|H_1| , |H_2| < 1 \, ,
\;\;\;\;\;
|H_1| , |H_2| > 1 \, .
\ee

The parametrizations of eqs.~(\ref{first}) and (\ref{second})
make explicit the factorization property of the manifold: to
make connection with the general parametrization of $SO(2,n)/
[SO(2) \times SO(n)]$ manifolds, we can make the field
redefinitions $(H_1=y_1 + i y_2, H_2 = y_1 - i y_2)$ and
$(T = \rho + \sigma, U= \rho - \sigma)$, which give
\be
Y^{(i)} = 1 - 2 \left( |y_1|^2 + |y_2|^2 \right)
+ | y_1^2 + y_2^2 |^2
\ee
and
\be
Y^{(ii)} = \left( \rho + \ov{\rho} \right)^2  -
\left( \sigma + \ov{\sigma} \right)^2 \, .
\ee

\newpage
\section*{References}
\begin{enumerate}
\item
\label{sergio}
S.~Ferrara, ed., {\em Supersymmetry} (North-Holland, Amsterdam,
1987).
\item
\label{noscale}
N.-P.~Chang, S.~Ouvry and X.~Wu, Phys. Rev. Lett.
51 (1983) 327;
\\
E.~Cremmer, S.~Ferrara, C.~Kounnas and D.V.~Nanopoulos,
Phys. Lett. B133 (1983) 61;
\\
J.~Ellis, A.B.~Lahanas, D.V.~Nanopoulos and K.~Tamvakis, Phys. Lett.
B134 (1984) 429;
\\
J.~Ellis, C.~Kounnas and D.V.~Nanopoulos,
Nucl. Phys. B241 (1984) 406 and B247 (1984) 373;
\\
S.~Ferrara and A.~Van Proeyen, Phys. Lett. B138 (1984) 77;
\\
U.~Ellwanger, N.~Dragon and M.~Schmidt, Nucl. Phys. B255 (1985) 549;
\\
M.~Dine, R.~Rohm, N.~Seiberg and E.~Witten, Phys. Lett. B156 (1985)
55;
\\
R.~Barbieri, S.~Ferrara and E.~Cremmer, Phys. Lett. B163 (1985) 143;
\\
C.~Kounnas and M.~Porrati, Phys. Lett. B191 (1987) 91;
\\
S.~Ferrara, C.~Kounnas, M.~Porrati and F.~Zwirner,
Nucl. Phys. B318 (1989) 75.
\item
\label{lhc}
S.~Ferrara, C.~Kounnas, M.~Porrati and F.~Zwirner,
Phys. Lett. B194 (1987) 366;
\\
C.~Kounnas, M.~Quir\'os and F.~Zwirner, Nucl. Phys. B302 (1988) 403;
\\
S.~Ferrara, C.~Kounnas and F.~Zwirner, Nucl. Phys. B429 (1994) 589;
\\
U.~Ellwanger, preprint LPTHE Orsay 94-106.
\item
\label{bz}
A.~Brignole and F.~Zwirner, Phys. Lett. B342 (1995) 117.
\item
\label{fayet}
P.~Fayet, Phys. Lett. 70B (1977) 461;
\\
R.~Casalbuoni, S.~De~Curtis, D.~Dominici, F.~Feruglio and R.~Gatto,
Phys. Rev. D39 (1989) 2281.
\item
\label{cdf}
L.~Castellani, R.~D'Auria and P.~Fr\'e, `Supergravity and
Superstrings:
a Geometric Perspective', World Scientific, Singapore, 1991, p.~942.
\item
\label{gpr}
A.~Giveon, M.~Porrati and E.~Rabinovici, Phys. Rep. 244 (1994) 77.
\item
\label{cv}
E.~Calabi and E.~Vesentini, Ann. Math. (N.Y.) 71 (1960) 472;
\\
S.~Ferrara, L.~Girardello, C.~Kounnas and M.~Porrati, Phys. Lett.
B192 (1987) 368.
\item
\label{fmp}
S.~Ferrara, A.~Masiero and M.~Porrati, Phys. Lett. B301 (1993) 358.
\item
\label{dg}
S.~Dimopoulos and H.~Georgi, Nucl. Phys. B193 (1981) 150.
\item
\label{mariano}
C.~Kounnas, J.~Leon and M.~Quir\'os, Phys. Lett. B129 (1983) 67;
\\
C.~Kounnas, D.V.~Nanopoulos and M.~Quir\'os, Phys. Lett. B129 (1983)
223.
\item
\label{npglo}
D.~Amati, K.~Konishi, Y.~Meurice, G.C.~Rossi and G.~Veneziano,
Phys. Rep. 162  (1988) 169;
\\
N.~Seiberg and E.~Witten, Nucl. Phys. B426 (1994) 19 and B431 (1994)
484;
\\
N.~Seiberg, Nucl. Phys. B435 (1995) 129.
\item
\label{neq2}
M.~Gunaydin, G.~Sierra and P.K.~Townsend, Phys. Lett. B133 (1983) 72
and B144 (1984) 41; Nucl. Phys. B242 (1984) 244, B253 (1985) 573 and
B272 (1986) 598; Class. Quant. Grav. 3 (1986) 763;
\\
E.~Cremmer, C.~Kounnas, A.~Van~Proeyen, J.P.~Derendinger, S.~Ferrara,
B.~de~Wit and L.~Girardello, Nucl. Phys. B250 (1985) 385;
\\
Yu.M. Zinoviev, preprint hep-th/9409199.
\item
\label{neq4}
M.~de Roo, Nucl. Phys. B255 (1985) 515 and Phys. Lett. B156 (1985)
331;
\\
E.~Bergshoeff, I.G.~Koh and E.~Sezgin, Phys. Lett. B155 (1985) 71;
\\
M.~de Roo and P.~Wagemans, Nucl. Phys. B262 (1985) 644
and Phys. Lett. B177 (1986) 352;
\\
P.~Wagemans, Phys. Lett. B206 (1988) 241;
\\
M.~Porrati and F.~Zwirner, Nucl. Phys. B326 (1989) 162;
\\
V.A.~Tsokur and Yu.M. Zinoviev, preprint hep-th/9411104.
\item
\label{neq8}
C.M.~Hull, Phys. Rev. D30 (1984) 760; Phys. Lett. B142 (1984) 39
and B148 (1984) 297.
\item
\label{ss}
C.~Kounnas and M.~Porrati, Nucl. Phys. B310 (1988) 355;
\\
S.~Ferrara, C.~Kounnas, M.~Porrati and F.~Zwirner,
Nucl. Phys. B318 (1989) 75;
\\
C.~Kounnas and B.~Rostand, Nucl. Phys. B341 (1990) 641;
\\
I.~Antoniadis, Phys. Lett. B246 (1990) 377;
\\
I.~Antoniadis and C.~Kounnas, Phys. Lett. B261 (1991) 369;
\\
I.~Antoniadis, C.~Mu\~noz and M.~Quir\'os, Nucl. Phys. B397 (1993)
515.
\item
\label{gp}
A.~Giveon and M.~Porrati, Phys. Lett. B246 (1990) 54 and Nucl. Phys.
B355 (1991) 422;
\\
S.~Ferrara, C.~Kounnas, D.~L\"ust and F.~Zwirner, Nucl. Phys. B365
(1991) 431.
\item
\label{ferstr}
H.~Kawai, D.C.~Lewellen and S.-H.H. Tye, Nucl. Phys. B288 (1987) 1;
\\
I.~Antoniadis, C.~Bachas and C.~Kounnas, Nucl. Phys. B289 (1987) 87.
\item
\label{flipped}
I.~Antoniadis, J.~Ellis, J.S.~Hagelin and D.V.~Nanopoulos,
Phys. Lett. B194 (1987) 231, B208 (1988) 209, B213 (1988) 562,
B231 (1989) 65.
\item
\label{fereff}
S.~Ferrara, L.~Girardello, C.~Kounnas and M.~Porrati, Phys. Lett.
B192 (1987) 368 and B194 (1987) 358;
\\
I.~Antoniadis, J.~Ellis, E.~Floratos, D.V.~Nanopoulos and T.~Tomaras,
Phys. Lett. B191 (1987) 96.
\end{enumerate}
\end{document}